\documentclass[twocolumn,showkeys,superscriptaddress]{revtex4}
\usepackage{graphicx}
\begin{document}

\title{Invariant correlational entropy as a signature of quantum
phase transitions in nuclei}

\author{Alexander Volya}
\affiliation{Physics Division, Argonne National Laboratory,
Argonne, Illinois 60439, USA}
\author{Vladimir Zelevinsky}
\affiliation{National Superconducting Cyclotron Laboratory and
Department of Physics and Astronomy, Michigan State University,
East Lansing, Michigan 48824, USA}

\date{\today}

\begin{abstract}
We study phase transformations in finite nuclei as a function of
interaction parameters. The signature of a transition is given by
invariant correlational entropy that reflects the sensitivity of
an individual many-body state to changes of external
parameters; peaks in this quantity indicate the critical regions.
This approach is able to reveal the pairing phase transition,
identify the isovector and isoscalar pairing regions and determine
the role of other interactions. We show the examples of the phase
diagram in the parameter space.
\end{abstract}
%\pacs{}
\maketitle

Recently, an appreciable effort has been applied to understand and
classify quantum phase transitions
\cite{sachdev99,borrmann01,mulken01}. Although the concept of a
phase transition is strictly applicable in the thermodynamic limit
only, there are numerous examples of structural changes in
mesoscopic systems, ranging from molecular clusters and
semiconductors to atomic nuclei or quark systems, under variation
of control parameters. Proper understanding of such transitions is
crucial for areas from cosmology and formation of the universe to
quantum computing and decoherence. The typical feature of phase
transitions in small systems is the absence of discontinuities in
the observables, and, therefore, difficulty for identification and
classifications of such transitions. The counterparts of phase
changes in small systems involve restructuring, critical
sensitivity to parameters, and chaotic large-scale fluctuations.
In this work we study mesoscopic phase transitions in atomic
nuclei using shell model interactions, which are known to
reproduce the low-lying states in selected nuclei with a
remarkable quality. The instrument we suggest for such studies can
be similarly applied to other finite quantum systems.

We identify the presence of a phase transformation as an
enhancement in the {\sl invariant correlational entropy} (ICE)
which was introduced in \cite{sokolov98}. ICE provides a measure
of sensitivity of a given state in the many-body system to
variations of external parameters. From earlier studies of bosonic
models the increase of ICE is known to be associated with critical
points \cite{cejnar01}. We do not consider here thermal phase
transitions for a system in a heat bath although it is possible,
see \cite{sokolov98} and \cite{cejnar01}, to introduce an
effective temperature as a measure of fluctuations in the
dynamical response to the change of parameters.

Following \cite{sokolov98}, we assume a Hamiltonian $H(\lambda)$
that depends on a parameter $\lambda ,$ so that, in an arbitrary
basis $|k\rangle ,$ any eigenstate $|\alpha(\lambda) \rangle$ of
$H(\lambda)$ can be decomposed as $|\alpha(\lambda)\rangle=\sum_k
C_k^\alpha(\lambda) |k\rangle\,.$ The ICE is then defined as
\begin{equation}
S^{\alpha}(\lambda)=-\text{Tr}\left \{\rho^\alpha \ln(\rho^\alpha)
\right \} ,                                          \label{1}
\end{equation}
where $\rho^{\alpha}$ is the density matrix of the state
$|\alpha\rangle$ in the basis $k$ averaged over a small region
${\lambda} \in [\lambda, \lambda+\delta ]$,
\begin{equation}
\rho^\alpha_{k k'}(\lambda)=\overline{C^\alpha_k C^{\alpha
*}_{k'}}.                                            \label{2}
\end{equation}
The discussion of various entropy-like quantities of individual
wave functions can be found in Ref. \cite{sokolov98}. ICE is
basis-independent von Neumann entropy that reflects the
correlations between the wave function components along the
evolution path determined by the parameter $\lambda$. It should
not be confused with basis-dependent Shannon {\sl information
entropy}
\begin{equation}
I^{\alpha}=-\sum_{k}|C^{\alpha}_{k}|^{2}\ln |C^{\alpha}_{k}|^{2}
                                                   \label{3}
\end{equation}
that was extensively used for studying the degree of complexity of
the state $|\alpha\rangle$ with respect to the given basis
$|k\rangle$ \cite{izrailev90,big} without taking into account
correlations between the amplitudes $C^{\alpha}_{k}$.

For a pure state, i.e. without averaging, the density matrix has a
single non-zero eigenvalue equal to one because of the
normalization ${\rm Tr}\,\rho=1$. This results in $S^\alpha=0.$
After averaging, we deal with a mixed state, and the eigenvalues
of $\rho$ deviate from the trivial limit. As shown in
\cite{sokolov98}, higher orders of perturbation theory bring in
new non-zero eigenvalues. Averaging over only two discrete points
$\lambda$ and $\lambda'$ leads to a factorized matrix
$\rho^{\alpha}$ with two non-zero eigenvalues
\begin{equation}
r^\alpha_{\pm}=\left (1\pm |\langle \alpha(\lambda) |
\alpha(\lambda') \rangle| \right )/2.            \label{twop}
\end{equation}
Now ICE can change from 0 (in the absence of any evolution of the
state) to the maximum value $\ln 2$ reached for the orthogonal
states in the representative points $\lambda$ and $\lambda'$.
Thus, ICE in a basis-independent way shows how ``quickly'' the
eigenstate $\alpha$ reorients due to sudden change of interaction
from $H(\lambda)$ to $H(\lambda').$ This makes ICE an ideal tool
for studying phase transitions.

We apply the ICE tool to spherical shell-model systems, where the
general Hamiltonian includes independent particle orbitals and
two-body residual interactions,
\begin{equation}
H=\sum_{\bf 1}\epsilon_{1} a^\dagger_{\bf 1} a_{\bf 1} +
\sum_{{ 1} {2} {3} {4} L \Lambda}
\,V^{({1} {2}; {3} {4})}_{L}\,\left(
P^{({1}\,{2})}_{L\,\Lambda}\right)^{\dagger} \,
P^{({3}\,{4})}_{L\,\Lambda}\,.
                                        \label{int:inth}
\end{equation}
Here pair creation and annihilation operators couple corresponding
single-particle operators to an appropriate angular momentum, i.e.
$P^{(1 2)}=(1+\delta_{12})^{-1/2}[a_{\bf 1} \times a_{\bf 2}]_{L
\Lambda}.$ The usual pairing interaction is given by the $L=0$
term. We suppress the isospin degree of freedom, but it is
straightforward to include it in direct analogy to angular
momentum.

We start with a simple model, where the pairing interaction is
known to result in a phase transition in the BCS approach. $N$
identical nucleons occupy two single-particle orbitals
$\epsilon_1=-\epsilon$ and $\epsilon_2=\epsilon$ of equal
spherical degeneracy $\Omega_1=\Omega_2=\Omega$. They interact via
off-diagonal pair transfers determined by a parameter $\lambda$ as
$V^{(11;22)}_{0}=V^{(22;11)}_{0}=-2\lambda/\Omega.$ Let the system
be half-occupied, $N=\Omega$, which sets the chemical potential
$\mu=0.$ It is known from the BCS theory that the pairing
interaction causes the normal to superconducting phase transition,
and, in the asymptotic limit of a large system, any weak but
attractive pairing is sufficient for creating the Cooper
instability. For a finite system the BCS solution is approximate.
Because of the finite single-particle level spacing at the Fermi
surface, the BCS phase transition takes place at the critical
strength value,
\begin{equation}
\lambda_c^2=\frac{16 \epsilon^2}{\Omega^2}.       \label{lc}
\end{equation}

\begin{figure}
\begin{center}
\includegraphics[width=7 cm]{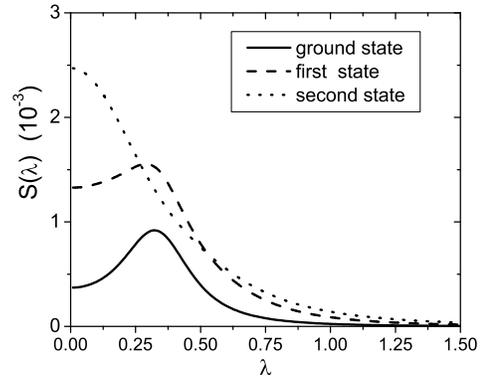}
\end{center}
\caption{Invariant correlational entropy of the ground and two
lowest pair-vibrational excited states calculated for the
two-level pairing model as a function of pairing strength
$\lambda$ and with averaging interval $\delta=0.01.$ \label{ipv}}
\end{figure}

The exact solution of the problem \cite{PRC65} determines the
ground state and the excited states of two types,
``pair-vibrational" states with seniority $s=0$ and redistribution
of the pairs, and states with broken pairs and $s\neq 0$. For each
state one can calculate ICE varying $\lambda$. The results for the
ground and two lowest pair-vibration states calculated with the
averaging interval $\delta=0.01$ are shown in Fig. \ref{ipv}. The
clear ICE peak in the ground state indicates the presence of the
phase transition. For $\epsilon=1$ and $\Omega=N=16$ the BCS
theory predicts a phase transition at $\lambda_c=0.25$.
Corrections to the BCS vacuum \cite{PRC65,rabhi02} make
$\lambda_c$ slightly larger, and ICE reaches its maximum at
$\lambda_c=0.3.$ In the excited states the phase transition is
quenched. The first excited state is collective and exhibits a
broadened peak of ICE. For the second excited state the peak
disappears although there is still a maximum of $S(\lambda)$ at
$\lambda=0$ indicating an enhanced sensitivity to switching on
pairing.

\begin{figure}
\begin{center}
\vskip 0.5 cm
\includegraphics[width=6 cm]{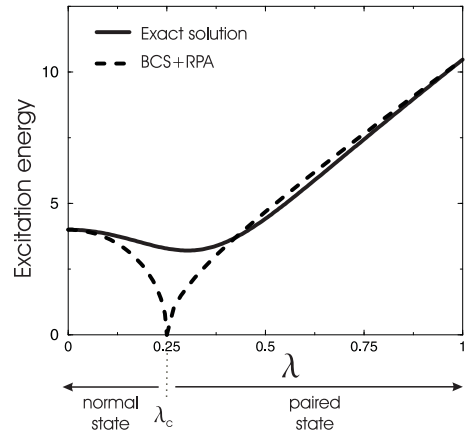}
\end{center}
\caption{Energy of the first pair-vibration level in the BCS+RPA
(dashed line with instability at $\lambda_{c}$) and exact solution
(solid line) that shows softening at the shifted point.
\label{two_levelBCS}}
\end{figure}

It is also possible to recognize the signature of the phase
transition in the exact energy spectrum. In Fig.
\ref{two_levelBCS} the excitation energy of the first pair
vibration state is shown. The softening of this mode at the same
point $\lambda_c=0.3$ is characteristic for the phase transition
and supports the level crossing interpretation. With the dashed
line we show the same state in the random phase approximation,
RPA, built on the BCS condensate \cite{PRC65}. The instability of
the BCS+RPA approach for a soft mode is an artifact of the invalid
approximation, see also \cite{rabhi02}.

\begin{figure}
\begin{center}
%\vskip 0.5 cm
%\epsfxsize=10.0cm \epsfbox{FIG1.EPS}
\includegraphics[width=7 cm]{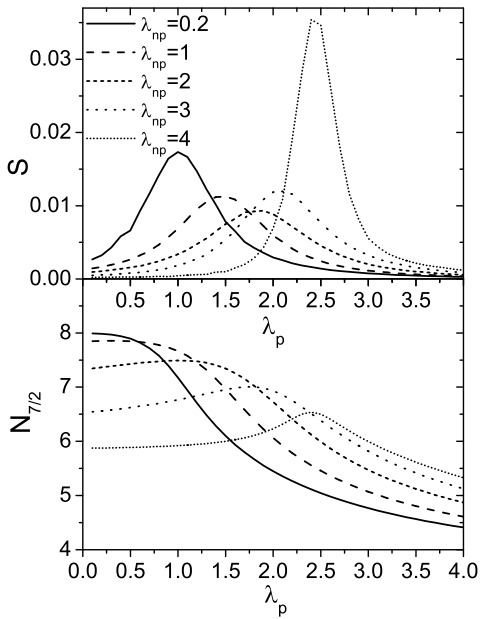}
\end{center}
\vskip -1 cm \caption{ICE (computed with averaging interval
$\delta=0.1$), upper panel, and occupancy of the lowest $f_{7/2}$
orbital, lower panel, as a function of pairing strength for
different scales of non-pairing interactions, $\lambda_{np}=$0.2,
1, 2, 3, and 4.  \label{ielca} }
\end{figure}

\begin{figure*}
\begin{center}
\vskip -3 cm
\includegraphics[width=15 cm]{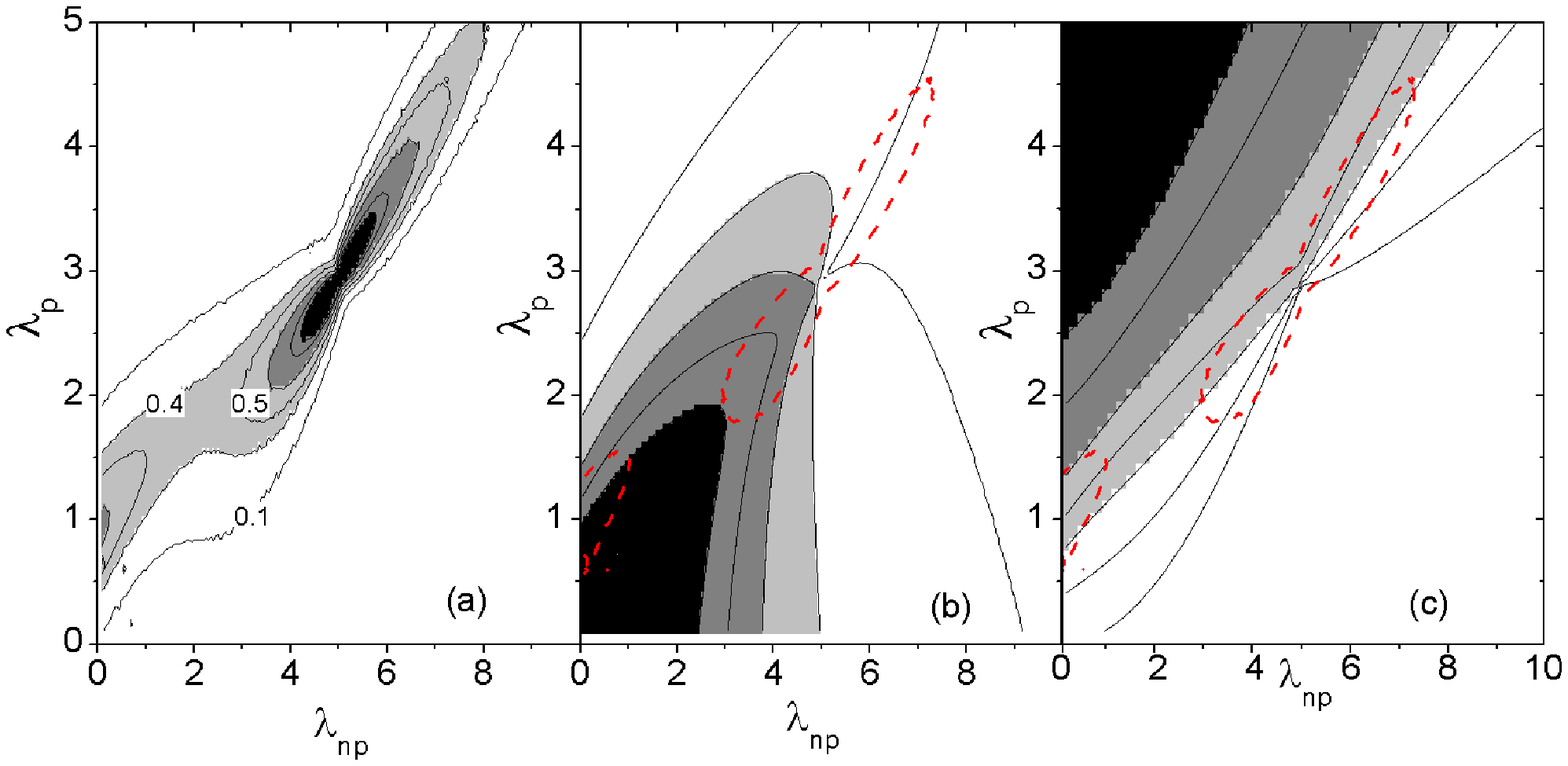}
\end{center}
\vskip -7.0 cm
\caption{\label{PDCa}Phase diagram of the shell model ground state
wave function of $^{48}$Ca. The parameters are the pairing
strength $\lambda_{p}$ and an overall strength $\lambda_{np}$ of
remaining non-pairing matrix elements. Panel (a) shows ICE with
the averaging interval $\delta=0.05$ in both scales. ICE is
normalized by a factor 0.015. Panels (b) and (c) show the
occupancy of the $f_{7/2}$ orbital and the average number of
correlated pairs ${\cal N}_1$, Eq. (\ref{fig3c}), respectively. To
emphasize the consistency between these quantities and location of
phase changes, the dotted contours in panels (b) and (c) outline
the region where ICE is enhanced above the value of 0.5.}
\end{figure*}

As a realistic example we consider the case of the $N=28$ subshell
closure in $^{48}$Ca. It has been emphasized in Ref.
\cite{volyaEP} that although the BCS approach predicts no paired
ground state in this nucleus, the exact solution indicates
presence of pairing correlations. The interplay of weak pairing
and self-consistent monopole renormalization of single-particle
energies leads to correlation energy of nearly 2 MeV. The
specifics of pairing correlations and self-consistent mean field
in $N=28$ nuclei have recently been a subject of experimental
\cite{cottle02} and theoretical \cite{werner96,lalazissis99}
interest. Here we go beyond the mean field and solve exactly the
many-body Hamiltonian with the effective interaction
\cite{richter91} in the plane of parameters $\lambda_p$ and
$\lambda_{np}$ that scale all pairing ($T=1$, $L=0$) and remaining
non-pairing matrix elements, respectively.

The results for the ground state of $^{48}$Ca are given in Figs.
\ref{ielca} and \ref{PDCa}. The peaks in ICE in the upper panel in
Fig. \ref{ielca} unambiguously identify the pairing phase
transition. The drop in the occupancy of the lowest $f_{7/2}$
orbital shown in lower panel is also consistent with this behavior
indicating the disappearance of subshell closure and a transition
into a paired state. Five curves in both panels of Fig.
\ref{ielca} allow one to track the evolution of the phase
transition as a function of strength $\lambda_{np}$ of non-pairing
matrix elements. Panel (a) in Fig. \ref{PDCa} shows ICE with a
contour plot in the $\lambda_{p}$ - $\lambda_{np}$ plane. Higher
values are indicated with shaded areas. ICE is enhanced along the
diagonal outlining, as in a phase diagram, the boundary between
the normal Fermi state (lower right) and superconducting paired
state (upper left). The panel (c) confirms this identification of
phases. Here the effective number of pairs ${\cal N}_{T=1}$,
computed as an expectation value of the operator \cite{big}
\begin{equation}
\hat{{\cal N}}_{T}=\sum_{1 2} \left (P^{(11)}_{L=1-T,\,T}\right
)^\dagger P^{(2 2)}_{L=1-T,\, T}            \label{fig3c}
\end{equation}
in the ground state, is shown. This quantity reaches its maximum
value ${\cal N}_1=(1/4)(N-s)(\Omega-N-s+2)$, where $\Omega$ is the
total capacity of the valence space, and seniority $s$ gives the
number of unpaired particles, in the degenerate limit, i.e. when
the pairing becomes so strong that differences in single-particle
energies can be ignored. The maximum number of pairs is expected
for a fully paired state with $s=0$, which for our model, with
$\Omega=20$ and $N=8$, leads to ${\cal N}_1=28$. The non-pairing
interactions create a random background, mixing seniorities and
adding some statistical number of pairs to any state. This
background can be estimated by averaging operator $\hat{{\cal
N}}_1$ over all many-body states, $\overline{{\cal N}_1}
=[N(N-1)]/[2 (\Omega-1)]$, which gives $\overline{{\cal N}_1}
\approx 1.5 $ in our model. Indeed, the major part of the valley
in Fig. \ref{PDCa}(c), shown as a white area, is levelled at
around this value. In the phase transition region, along with the
enhancement of ICE, the number of correlated pairs quickly rises,
finally becoming close to 28. The full saturation at ${\cal N}_1
=28$ is not expected to be reached, since it is only possible for
the constant pairing interaction.

The enhancement of ICE not only allows one to localize the phase
transition but also quantifies the sharpness of the transformation
and the size of the critical region. In macroscopic BCS theory it
is usually assumed that the non-pairing interactions renormalize
quasiparticles but  essentially do not participate in the phase
transition. The shell model analysis \cite{big} demonstrated that
this is not the case, at least in a finite system. As seen from
Figs. \ref{ielca} and \ref{PDCa}, the scaling factor
$\lambda_{np}$ plays a significant role in the overall picture. At
$\lambda_{np}=0$, the pairing phase transition is quite sharp and
takes place slightly below $\lambda_{p}=1.0.$ Indeed, the
comparison of earlier studies \cite{volyaEP, belyaev} shows that
pairing would be stable and treatable with the BCS if no other
interactions had been present in $^{48}$Ca. Presence of other
parts of residual interaction softens and widens the transitional
region, putting the physical state $\lambda_{np}=\lambda_{p}=1$ in
the normal domain below the phase transition peak, although within
the region of enhanced pair fluctuations. It was also shown
earlier \cite{big} that other interactions destroy the purity of
the seniority classes (in the first order the ground state
acquires the contribution of $s=4$) moving the dynamics to
many-body quantum chaos \cite{big,PRC65}.

The non-pairing interactions shape the properties of the mean
field and redefine effective single-particle energies; on the
other hand, they induce chaotic residual dynamics. In Fig.
\ref{PDCa}(b), the occupancy of the $f_{7/2}$ orbital is mapped in
the same parameter space. The lowest $f_{7/2}$ level in $^{48}$Ca
is separated from other single-particle levels by almost 2 MeV. At
weak residual interactions the ground state is the slightly
perturbed Fermi sea, with nearly all $N=8$ valence neutrons
occupying the $f_{7/2}$ level. This region is seen as the dark
area in the lower left corner in Fig. \ref{PDCa}(b). As
$\lambda_{p}$ grows, the pairing phase transition breaks the
magicity, and the depopulation of the $f_{7/2}$ orbital is
consistent with the line of pairing phase transition, see also
Fig. \ref{ielca}. Along the $\lambda_{np}$ axis, the original
state with the fully occupied lowest orbital also disappears
around $\lambda_{np}=4$, similarly to the change of magic numbers
far from stability. The bare single-particle energies correspond
well to the mean field defined with respect to the $^{40}$Ca core.
The increase of residual interactions renormalizes the mean field
and at some point makes the original single-particle basis
inconsistent with the potential. Strong residual interactions make
the dynamics chaotic that can be interpreted as a non-zero
single-particle temperature \cite{big,PRC65}. The occupation of the lowest
orbital falls from $N_{7/2}=8$ to the ``thermal'' occupation
$N_{7/2}=8(N/\Omega)=3.2.$ When non-pairing interactions grow to
around $\lambda_{np}=5$ and beyond, the pairing to normal phase
transition again sharpens. Here the rise in ICE is much higher
than at $\lambda_{np}=0$. In order to create a coherent paired
state, pairing forces need to overcome not only the spread of
single-particle energies but also random motion generated by
strong residual interactions.

From ICE in Fig. \ref{PDCa}(a) we learn that the evolution of the
wave function driven by increase of non-pairing interactions is
not followed by an enhancement of entropy. Since with a closed
proton core a non-perturbative change in the mean field, such as
onset of deformation, is not expected, the evolution of spherical
orbitals, as well as chaotization of dynamics or rise in effective
temperature, are not associated with any phase transition. The
role of $\lambda_{np}$ and the mean field in the pairing phase
transition in nevertheless eminent. Supporting earlier studies in
\cite{big,harting00,egido00}, the sharpness of change and the
transitional region between normal and paired states are
influenced significantly by non-pairing interactions.

\begin{figure}
\begin{center}
%\vskip 0.5 cm
%\epsfxsize=10.0cm \epsfbox{FIG1.EPS}
\includegraphics[width=7 cm]{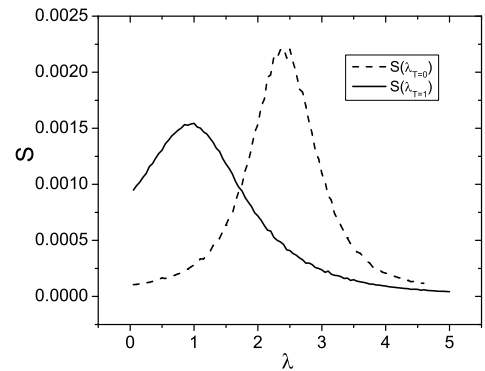}
\vskip -1 cm
\end{center}
\caption{The evolution of ICE: along the line of increasing
isovector pairing $\lambda_{T=1}$ at fixed
$\lambda_{T=0}=\lambda_{np}=1$,solid line; as a function of
$\lambda_{T=0}$ at fixed $\lambda_{T=1}=\lambda_{np}=1$, dashed
line. Interval $\delta=0.05$ was used here. \label{ielmg} }

\end{figure}
\begin{figure*}
\begin{center}
\vskip -3 cm
\includegraphics[width=15 cm]{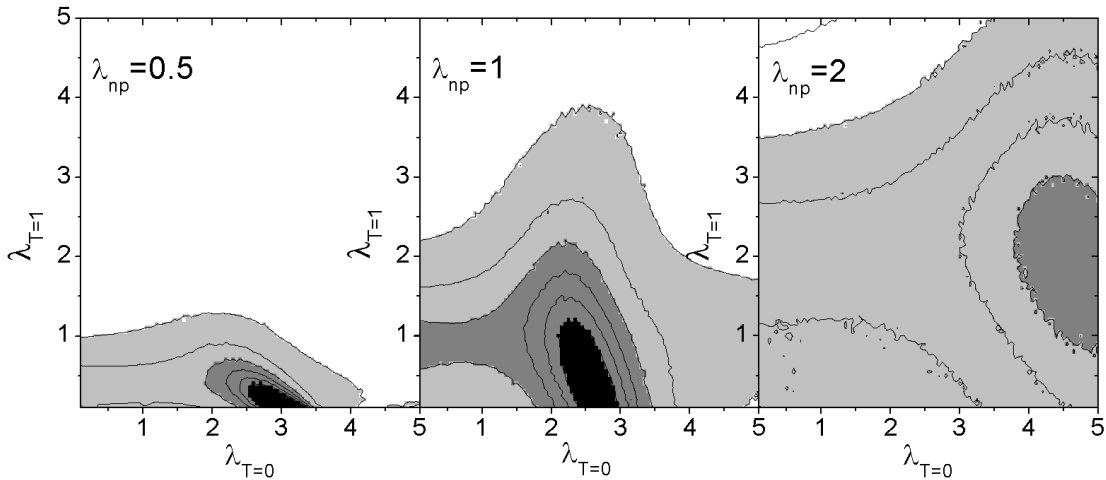}
\end{center}
\vskip -7.5 cm
\caption{ICE of the ground state in $^{24}$Mg computed on the
plane $\lambda_{T=0} - \lambda_{T=1}$. The three panels show the
evolution of the phase diagram depending on the overall scale of
non-pairing matrix elements $\lambda_{np}.$ The averaging interval
of $\delta=0.05$ was used, and the entropy is scaled (divided) by
0.015. \label{PDMg}}
\end{figure*}

\begin{figure}
\begin{center}
\vskip -0.2 cm
\includegraphics[width=7 cm]{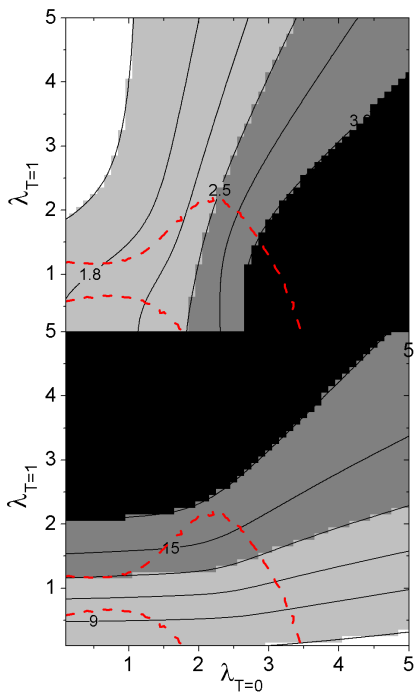}
\end{center}
\vskip -0.5 cm
\caption{Effective average number of $T=0$ and $T=1$ pairs, upper and
lower panels, respectively, for the ground state of $^{24}$Mg at
$\lambda_{np}=1$ on the $\lambda_{T=0} - \lambda_{T=1}$ plane. The
dashed line outlines the regions where ICE is enhanced above 0.45.
\label{PPMg}}
\end{figure}

As our last example, we consider the $sd$ shell model for
$^{24}$Mg. At $N=Z$, the competition of isovector, isoscalar,
deformation and alpha-particle correlations makes the physical
picture very complex. Empirical information \cite{macchiavelli00},
studies of simple models
\cite{evans81,dussel86,engel97,baldini03,palchikov01} and analytic
works \cite{ropke98}, supplemented by the direct shell model
diagonalization \cite{vogel99}, support the plausible existence of
different phases. However, it still remains unclear whether these
phases are pure enough to be identified, or whether they are
separated by distinct critical regions. Here the ICE provides a
helpful tool.

For this example we extend the list of parameters to include
$\lambda_{T=1}$, $\lambda_{T=0}$ and $\lambda_{np}$, which scale
isovector ($T=1$, $L=0$), isoscalar ($T=0$, $L=1$), and all
remaining two-body matrix elements in the Hamiltonian,
respectively. The shell model is defined in this example with
single-particle energies and interaction matrix elements from Ref.
\cite{brown88}. In  Fig. \ref{ielmg}, where ICE is shown as a
function of $\lambda_{T=1}$ with $\lambda_{T=0}=\lambda_{np}=1$ as
a solid line and as a function of $\lambda_{T=0}$, while
$\lambda_{T=1}=\lambda_{np}=1$, as a dashed line, indicates again
a phase transition. The contour plot of ICE for this example is
shown in Fig. \ref{PDMg}. We limit ourselves here to three
different strengths of non-pairing interactions $\lambda_{np}=0.5,
1,$ and 2. This already gives a general idea of the ICE evolution.
The valley in the vicinity of small $\lambda_{T=0}$ and
$\lambda_{T=1}$ that corresponds to a normal state is clearly
separated, and with increase of the strength of non-pairing
interactions this region widens. However, at large $\lambda_{np}$
the sharpness of the phase transition is significantly reduced;
note that the scale in the right panel in Fig. \ref{PDMg} is ten
times smaller. This effect is analogous to the situation observed
in $^{48}$Ca, where, except for some special area, non-pairing
interactions smeared the phase transition region.

The only valley clearly separated by the mountains of ICE
corresponds to the normal state. To identify phases of $T=1$ and
$T=0$ pairing, similarly to our previous example, we calculate the
number of coherent pairs for $\lambda_{np}=1$ case shown in Fig.
\ref{PPMg}. We consider isovector, $T=1,\,L=0$, and isoscalar,
$T=0,\,L=1$, pairs (\ref{fig3c}). For both limiting cases of pure
isovector and isoscalar pairing, ${\cal N}_{T}$ can be calculated
using group algebra, see for example \cite{engel98}. In the states
of degenerate isovector pairing model
\begin{equation}
{\cal N}_1 =\frac{1}{2} \left[ \frac{1}{4}(N-s)(\Omega-N-s+6)+{\bf
t}^2-{\bf T}^2\right ]\,,                        \label{N1}
\end{equation}
where ${\bf t}$ is isospin of unpaired particles, while ${\bf T}$
is total isospin of the system. Thus, in the state with maximum
isovector pairing, $s=0$, ${\bf t}={\bf T}=0$, one would expect
${\cal N}_1=N(\Omega-N+6)/8$, which in our case of $N=8$ and
$\Omega=24$ leads to ${\cal N}_1=22.$ Interchanging angular
momentum and isospin and taking normalization of operators into
account we obtain ${\cal N}_0=N(\Omega-N+6)/(2\Omega)$, which
results in maximum ${\cal N}_0\approx 3.7,$ whereas statistically
$\overline{{\cal N}_1} \approx 1.8$ and $\overline{{\cal
N}_0}\approx 0.3.$ This analysis along with Fig. \ref{PPMg} allow
us to confirm the presence of isovector and isoscalar paired
states in the upper left and lower right areas of the phase
diagram. The isovector pairing phase appears as a shaded area in
the lower panel of Fig. \ref{PPMg}, and the shaded area of the
upper panel indicates the isoscalar pairing state. Furthermore,
the enhancement of ICE, the region outlined by a dashed line on
both panels of Fig. \ref{PPMg}, is consistent with the presence of
the phase transition from normal to either of superconducting
phases; this is the area of the most rapid rise in the number of
coherent pairs ${\cal N}_0$ or ${\cal N}_1.$

Further examination of Fig. \ref{PDMg} indicates a notable absence
of any additional phase transition, as for example between
isovector-isoscalar and possibly alpha-clustering phases. Clearly
there are no phase separating lines far from the origin, i.e. at
large $\lambda_{T=0}$ and $\lambda_{T=1}.$ Thus, a continuous path
between isovector and isoscalar phases does not involve a
transitional behavior, unlike a transition from the normal state
to any paired state.

The point corresponding to the realistic nucleus, $\lambda_{T=1}=
\lambda_{T=0}=\lambda_{np}=1$, is located in the transitional
region from the normal to isovector pairing phase on the side of
the paired state. On the other hand, the isoscalar pairing matrix
elements are to be increased by a factor of three to bring the
system onto a border between the normal phase and quasideuteron
$T=0$ pairing coherence, this seen best from Fig. \ref{ielmg}. Our
result, that in the $N=Z$ nucleus the $T=1$ pair coherence is
enhanced in contrast to $T=0$, is supported by other findings
\cite{macchiavelli00,vogel99}.

Let us summarize the results for our shell model examples. In
$^{48}$Ca we are able to unambiguously identify the pairing phase
transition. This nucleus would be in the paired state if no
interactions of non-pairing type had been present. The incoherent
residual interactions in reality put the ground state below the
phase transition line although in the region of large fluctuations
preceding the onset of well developed pairing. This explains
earlier results \cite{volyaEP}: the failure of the BCS theory and
large pairing correlation energy found in the exact solution. We
have also found that the change in non-pairing interactions does
not lead to a phase transition, however it influences
substantially the normal-to-pairing critical region. Weak residual
incoherent interactions smear the phase transition as they
facilitate the pair excitation. Strong non-pairing interactions
create complexity in dynamics, and a strong phase transition
occurs when such system is ``cooled'' and the superconducting
order is established.

Considering the $^{24}$Mg nucleus we observed a clear phase
transition between the normal and superconducting state. We
identified the regions where two different types of pairing,
isovector and isoscalar, exist. The $^{24}$Mg nucleus is found, in
agreement with earlier results, to be in the isovector phase and
within the critical region of the normal to superconducting
transition. The isoscalar pairing should have been three times
stronger compared to the realistic shell model forces in order to
lead to the corresponding condensate. Within limited shell model space
we found no evidence of a
phase transition between different types of pairing or
alpha-clustering.

To conclude, in this work we suggested a new theoretical technique
for tracking and quantitatively studying the phase transitions in
nuclei and, more generally, in mesoscopic systems. Invariant
correlational entropy serves not only as an indicator of the
degree of complexity along the spectrum \cite{big} but as a
powerful tool for identifying the regions of rapid changes of
structure in response to variations in a control parameter, here
it was a strength of the interaction. A mesoscopic system is
sufficiently large to display statistical regularities but still
sufficiently small to allow one to probe, theoretically and
experimentally, individual wave functions. Such a system
significantly alters the features of phase transitions, and opens
a door for ``secondary" interactions to leave a noticeable
footprint. The ICE clearly shows the sensitive spots of the
parameter space.

\begin{acknowledgments}
The authors appreciate useful comments of R. Chasman and K.
Langanke and acknowledge support from the U.S. NSF, grants
PHY-0070911 and PHY-0244453, and from the U. S. Department of
Energy, Nuclear Physics Division, under contract No.
W-31-109-ENG-38.

\end{acknowledgments}
%\bibliographystyle{elsart-num1}
%\bibliography{pairing,phase,volya,shellmodel,vz,random}

\end{document}